\documentstyle[12pt]{article}
\newcommand{\cA}{\cal{A}}
\newcommand{\cO}{\cal{O}}

\begin{document}
\begin{flushright}
  { \large \rm NTZ 24/96 }
\end{flushright}
\vspace{1cm}
\begin{center}
{\LARGE\bf
Perturbative Regge asymptotics}
\end{center}
\begin{center}
{\LARGE\bf
       in the case of non-vacuum exchange \footnote{Contribution to the
Zeuthen Workshop on QCD and QED at higher order, Rheinsberg, April
1996.} }
\end{center}
\vspace{1cm}
 \begin{center}
{\large\bf R. Kirschner}\\
{\sl Institut f\"ur Theoretische Physik,
        Universit\"at Leipzig, \\
        D-04109 Leipzig, Germany}
\end{center}
\vspace*{3.0cm}
\begin{abstract}
Some results on the perturbative Regge asymptotics are reviewed. The
concepts of the reggeon interaction approach and the double logarithmic
approximation are outlined.
\end{abstract}

\newpage

\section{Results on non-vacuum exchange}

The leading logarithmic ($\ln s$ ) and the double-logarithmic
approximation schemes provide results about the perturbative Regge
asymptotics. These results apply to semi-hard processes where besides of
the large energy $s$ there is a second large scale $Q^2$ controlling the
coupling, $ s \gg Q^2 \gg \Lambda_{QCD}^2 $. The predictions for the
small $x$ behaviour of structure functions and on features of the
hadronic final state in deep-inelastic scattering at small $x$ are
 in qualitative agreement with experiment and usually attention is
paid to the features deviating from the GLAP evolution. On the other
hand the $Q^2$ dependence of the small $x$ asymptotic results is
determined by the asymptotics of the anomalous dimensions for the
continued moment number $j$ (angular momentum) approaching some integer
value. The relation of the perturbative Regge results to
 anomalous dimensions is well known \cite{Ja,KL82}.   These
asymptotic anomalous dimensions involving
the contributions from all orders are useful for checking complete
higher
order calculations and for improving the kernels in the GLAP evolution
\cite{CCH,BV}.

The perturbative Regge asymptotics for the vacuum exchange channel, the
BFKL pomeron \cite{BFKL}, has now its place in semi-hard
phenomeno\-logy.
It is natural to look for situations where the asymptotics of other
exchange channels shows up.

The case of flavour non-singlet exchange do\-mi\-nated by fermion and
antifermion with the same chirality in the $t$ channel has been
stu\-died
for many years. The first methods of double-logarithmic approximation to
the perturbative Regge asymptotics  have been worked out in QED
\cite{GGLF} and partially extended to QCD \cite{RK81}. The method of
separation of the softest particle \cite{KL82} provided an
essential simplification and allowed to treat both signature channels.

Alternatively the two-fermion exchange can be treated in close analogy
to
the two-gluon exchange of the BFKL pomeron \cite{RK94}. In this way all
contributions with $ \ln s $ from the longitudinal part of each loop
integral are summed up. There may be further single $ \ln s $
contributions. In this approach the negative signature case has not
been treated yet, because it is related to a three reggeon exchange.

The positive signature part of this channel contributes to the flavour
non-singlet $F_2 (x,Q^2) $ structure function \cite{KL82}.
The application of the negative signature part to the flavour
non-singlet part of the spin structure function $ g_1(x, Q^2) $ has been
considered first in \cite{BER95}.

The flavour singlet part of $g_1 (x,Q^2) $ cor\-res\-ponds to odd parity
exchange. Besides of the quarks also two gluons can be exchanged.
However at least one of the two gluons is of a different type (helicity
state) compared to the one gene\-rating the asymptotics in the vacuum
(BFKL) channel \cite{FK}.  The leading double log asymptotics has been
worked out recently \cite{BER96}. Both quark-antiquark and two-gluon
intermediate reggeon states give rise to a singularity near $j= 0$ and
the mixing of these channels has to be considered.

The positive signature part of the odd parity channel can be related to
the spin counterpart of the structure function $F_3 (x, Q^2)  $, which
is called $g_5 $ \cite{BK} or $a_1 $\cite{MESS}.

Odd parity can in principle be transferred by more than two leading
(BFKL type) gluons. The corresponding coupling structure (impact factor)
 is not known and is eventually related to higher twist. This would
give rise to a singularity near $j = 1$ beyond the leading logarithmic
approximation.

From studying the multi-Regge effective action \cite{KLS} it is evident
that there are two types of fermion exchanges characterized by chirality
or helicity and resulting in the same asymptotics.  The resulting two
kernels of reggeized two-fermion interactions have been treated in
analogy to the BFKL case \cite{RK94}. The case with equal chira\-lity of
the two fermions has been mentioned above. The other case of opposite
chirality (parallel helicity) of the exchanged quark and antiquark
shows some unusual features \cite{KMSS}. The leading Regge asymptotics
(near $j = 0 $)  has no double logarithmic contributions. Double
logs contribute at the subleading ( $j = -1 $) level and are
connected  to the (one - loop) GLAP evolution.
These features are expected to show up in the structure function $h_1
(x, Q^2)$ which appears in the Drell-Yan process and measures the
transversity distribution \cite{RSJ}.

\section{The reggeon interaction approach}

The multi-Regge effective action is a convenient formulation of the
leading $\ln s$ approximation, i.e. for summing contributions with the
leading logarithm arising from the longitudinal momentum integration. It
has been proposed \cite{L91}  as a starting point for investigating
unitarity corrections (multi-reggeon exchange with multi-Regge
kinematics in all $s$-channels)  and genuine next-to-leading
logarithmic corrections (quasi-multi-Regge kinematics).
This action can be derived from the QCD action by separating the gluon
and the quark fields into modes according to the multi-Regge kinematics.
The "heavy" modes, which do not correspond to exchanged or scattered
particles, are eliminated. The exchanged modes correspond to momenta $k$
with
\begin{equation}
k_+ k_- \ll \vert \kappa^2 \vert,
\end{equation}
where $k_{\pm} $ are the light cone components (essentially in the
direction of the incoming particles) and $\kappa$ are the transverse
components. In the light-cone axial gauge the transverse components of
the vector potential $A_{\sigma }, \sigma = 1,2 $ describes the
physical degrees of freedom of the gluon. There are two types of
exchange fields expressed in terms of the modes (1) of $A_{\sigma}$,
\begin{eqnarray}
{\cA}_+ = \partial_-^{-1}  \partial_{\sigma} A_{\sigma}  \ \ \  ({\rm
G} ),  \cr
{\cA}^{\prime } = i \epsilon_{\rho \sigma} \partial_{\sigma} A_{\rho}
\ \ \  ({\rm G}^{\perp} ).
\end{eqnarray}
The first exchange field gives rise to the leading asymptotics
(${\cO}(s^{1+ \omega}), \omega = {\cO}(g^2)$), whereas the exchange of
each ${\cA}^{\prime}$ leads to a suppression by one factor of $ s^{-1}$.
There are also two types of fermion exchanges characterized by the two
chirality projections. Each of them leads to a suppression by a factor
of $s^{-1/2}$ compared to the leading asymptotics. The two gluon and the
two fermion exchanges can be related by supersymmetry.

The terms in the effective action describing the non-leading
${\cA}^{\prime}$ exchange are not completely known yet.
The effective production vertices with the other three exchanges give
rise to the following two-reggeon interaction kernels \cite{RK94},
\begin{eqnarray}
K_{GG} &= \vert \kappa_1 - \kappa_1^{\prime } \vert^{\-2}
\left ( {\kappa_1 \kappa_2^* \over
 \kappa_1^{\prime } \kappa_2^{\prime *} } + {\rm c.c.} \right ), \cr
K_{F \overline F} &= \vert \kappa_1 - \kappa_1^{\prime } \vert^{-2}
\left ( \vert {\kappa_1  \over
 \kappa_1^{\prime } }\vert^{\omega } +
\vert {\kappa_1  \over
 \kappa_1^{\prime } }\vert^{- \omega }
 {\kappa_1^* \kappa_2 \over
 \kappa_1^{\prime *} \kappa_2^{\prime } }  \right ), \cr
K_{FF} &=
 \vert \kappa_1 - \kappa_1^{\prime } \vert^{-2}
\left ( { \kappa_2^* \over
 \kappa_2^{\prime *}  } +
 {\kappa_1^* \over
 \kappa_1^{\prime* }  } \right ).
\end{eqnarray}
The kernel $K_{GF}$ is not needed for our discussion here. The first
expression in (3) is the well known BFKL kernel \cite{BFKL} in the
complex notation
for the transverse momentum vectors. The momentum transfer is $\kappa_1
+ \kappa_2 = \kappa_1^{\prime } + \kappa_2^{\prime } $.
The second kernel $K_{F \overline F}$  corresponds to  the equal
chirality exchange. The $\omega$-dependent factors appear due to the
double logarithmic behaviour at $\kappa \ll \kappa^{\prime }$ and at
$\kappa \gg \kappa^{\prime }  $. The third kernel describes opposite
chirality
exchange and does not lead to a logarithmic transverse momentum
integral.

Taking the reggeization into account the kernels are modified and become
infrared finite ope\-ra\-tors,
\begin{equation}
H_{ij} = K_{ij} - \alpha_i (\kappa_1) \delta(\kappa_1 -
\kappa_1^{\prime } )
 - \alpha_j (\kappa_2) \delta(\kappa_2 -
\kappa_2^{\prime } ).
\end{equation}
$\alpha_i (\kappa )$ is the one-loop trajectory function of the gluon
($i = G$ ) or the quark ($i =  F, \overline F$). The equations and the
resulting $2 \rightarrow 2$ reggeon Green functions are analogous to the
BFKL case. We restrict ourselves to the case of vanishing momentum
transfer ($\kappa_1 = - \kappa_2 = \kappa $).

\begin{eqnarray}
f^{(ij)} (\omega, \kappa, \overline \kappa ) =   \cr
{1 \over 2 \pi^2} \sum_{r = -\infty}^{\infty}
\int_{-\infty}^{\infty} d\nu
{ f_{n,\nu}^{ (ij) *} (\kappa)
f_{n,\nu}^{ (ij) } (\overline \kappa)  \over
\omega - {g^2 C^{(ij)} \over 8 \pi^2 }
\Omega^{(ij)} (n,\nu, \omega ) }.
\end{eqnarray}
$\Omega^{(ij)} (n,\nu, \omega)$ are the eigenvalues of the kernel
$H_{ij}$ and $ f_{n,\nu}^{(ij)}$ are the corresponding eigenfunctions
\cite{RK94},
\begin{equation}
 f_{n,\nu}^{ (ij) } (\kappa) =  (d_i (\kappa) d_j (\kappa) )^{-1/2}
\vert \kappa \vert^{-1/2 + i\nu} \left ({\kappa \over \vert \kappa \vert
}\right )^{n}.
\end{equation}
$d_i (\kappa)$ denotes the propagator of the exchanged quark ($ i = F,
\overline F$) or gluon ($i = G $).

The Regge singularities (branch points) are located at the point in
$\omega$ where two poles in $\nu$ arising from the denominator pinch the
$\nu$ integration contour.
Also the resummed anomalous dimensions $\gamma (\omega)$ are obtained
from this denominator. It is given by a zero $\nu (\omega )$ such that
$\gamma (\omega) = \delta + i \nu[\omega ), \delta = 1/2 $ for GG or
$\delta = 0 $ for $F \overline F$
has a perturbative expansion, i.e. vanishes for $g^2 \rightarrow 0$
\cite{Ja}.

The analogy to the BFKL case extends also to the remarkable properties
of conformal symmetry, factorizability and integrability.

\section{The double logarithmic approximation}

Double logarithmic contributions arise in different situations and
kinematical regions. The small $x$ asymptotics of the GLAP evolution
matches the large $ \kappa^2$ asymptotics of the BFKL evolution. Both
are
dominated by double lo\-ga\-rithms from strong ordering in logitudinal
and transverse momenta. The Regge asymptotics in the channels of equal
chirality quark - antiquark exchange ($F \overline F$) and odd parity
two-gluon exchange ($G G^{\perp}$)  is dominated by double logs, whereby
the transverse momentum integral is logarithmic in two regions (compare
$K_{F \overline F}$ in (3) ). The GLAP evolution in these channels at
small $x$ accounts for only the logarithms from strongly ordered
transverse momenta and therefore does not match the perturbative Regge
behaviour \cite{EMR}.

Summing all double logs in these channels leads
to involved linear integral equations in momentum representation. The
method of separation of the
softest particle leads directly to much simpler equations in terms of
partial waves with a quadratic non-linearity \cite{KL82}.
One separates a loop in the configuation of  transverse momenta in all
loops where the one of the separated loop is the smallest. If the loop
is built by a bremsstrahlung gluon Gribov's theorem allows to express
the low $\kappa$ contribution of this loop by contributions with the
gluon radiated and absorbed by the external legs of the amplitude
without this soft loop $M(s, \kappa)$. The second argument indicates
that the transverse momenta in all loops are larger than $\kappa$.
The bremsstrahlung contribution has the form
\begin{equation}
\int_{\mu^2}^{s} {d\kappa^2 \over \kappa^2} \left (
\hat m_s \ln {-s \over \vert \kappa^2 \vert} +
\hat m_u \ln {s \over \vert \kappa^2 \vert} \right )
M (s, \kappa).
\end{equation}
The gluon loop leads in particular to a change of the colour state in
the $t$-channel. This is described by the matrices $\hat m_s$ and $\hat
m_u $ depending to which external legs the gluon is attached.
We change to patial waves
\begin{equation}
M^{\pm} (s, \kappa) = \int_{-i \infty}^{i \infty} { d\omega \over 2 \pi
i }  \left (s \over \vert \kappa^2 \vert \right )^{\omega}
\zeta^{\pm} (\omega) f^{\pm} (\omega) .
\end{equation}
The signature factor $\zeta^{\pm} (\omega) $ behaves for small $\omega $
as $\zeta^{+}(\omega ) \rightarrow 1,
\zeta^{-}(\omega ) \sim \omega $.
The positive signature contribution of the bremsstrahlung gluons to the
partial wave is given by
\begin{equation}
(\hat m_s + \hat m_u) {1 \over \omega} {d \over d\omega } f^{(+)}
(\omega )
\end{equation}
and the negative signature contribution by
\begin{eqnarray}
(\hat m_s + \hat m_u) {1 \over \omega^2} {d \over d\omega }(\omega
f^{(-)} (\omega ) )                               \cr
- (\hat m_s - \hat m_u) {1 \over \omega^2}
f^{(+)} (\omega ) ).
\end{eqnarray}
$\hat m_s + \hat m_u $ is diagonal and the element corresponding to the
colour singlet channel vanishes.  The positive signature part
contributes as a inhomogeneous term  to the negative signature channels,
which follows from the small $\omega$ behaviour of the signature factor.

The separated soft loop can also be built by a two-particle ($F
\overline F$ or $G G^{\perp}$) intermediate state in the $t$-channel.
This contribution to the amplitude is expressed by a loop integral
involving two amplitudes of lower order. In terms of partial waves we
obtain
\begin{equation}
{1 \over \omega } f^2 (\omega).
\end{equation}
This term conserves signature and all other $t$-channel quantum numbers.

It is easy to understand without explicite calculation that the exchange
$GG$ of two leading (BFKL type) gluons ${\cA}_+$ does not give rise to a
double log contributions with these gluons carrying the smallest
$\kappa$. By gauge invariance the amplitudes with external
${\cA}_+$ vanish proportional to the small transverse momentum of this
gluon. Therefore there is no logarithm from the small transverse
momentum integration range.

Now the Regge amplitude with the double logarithms of the types (7) and
(11)
 summed up is the solution of the equation in Fig. 1. Starting from the
Born term one iterates by adding bremsstrahlung loops and by inserting
soft $t$-channel intermediate states. The Born term has the form
${g^2 \over \omega} a_i$.  The coefficient $a_i$ depends on the colour
state in the $t$-channel. For the colour singlet state we have $a_0 =
C_2 $ for $F \overline F$ and $a_0 = 4 N$ for $G G^{\perp}$ exchanges.
In the parity odd flavour singlet case one has to consider the coupled
channel of both exchanges with Born terms describing the transitions in
the $t$-channel
$F \overline F \rightarrow F \overline F,
F \overline F \rightarrow G G^{\perp},
G G^{\perp} \rightarrow F \overline F,
G G^{\perp}  \rightarrow G G^{\perp}.$
The coefficients $a_0$ coincide with the residues at $j = 0$ of the
corresponding one-loop anomalous dimensions.

Due to the mentioned structure of $\hat m_s + \hat m_u $the derivative
terms (9) or (10) are absent in the colour singlet equation and we
have
just  quadratic equations. We write the  solutions for the $F
\overline F$ case.
\begin{eqnarray}
f_0^{(+)} (\omega) = 4 \pi \omega \left ( 1 -
\sqrt{1 - {g^2 C_2 \over 2 \pi^2 \omega^2 } } \right ), \cr
f_0^{(-)} (\omega) = 4 \pi \omega \cr \left ( 1 -
\sqrt{1 - {g^2 C_2 \over 2 \pi^2 \omega^2 }(1 - {1 \over 2 \pi^2 \omega}
f_V^{(+)} (\omega)      } \right ).
\end{eqnarray}
The solution for the colour octet channel $f_V^{(+)} (\omega )$ enters
the solution for the negative signature colour singlet partial wave. The
corresponding equation is of Riccati type and the solution is expressed
in terms of the parabolic cylinder functions.

The equation for the vitual Compton amplitude reduces to a linear
equation with the (appropriate projection of the) Regge amplitude as the
Born term. Therefore the large $Q^2$ behaviour is determined by
$f_0^{\pm} (\omega )$ (12), which turn out to be (up to a factor ( $8
\pi^2)^{-1}$) the asymptotic anomalous dimensions at $j = \omega
\rightarrow 0$.

Notice that the Born coefficient $a_0$ in the $G G^{\perp }$ channel is
large
compared to the $F \overline F$ channel. This implies for the position
of the Regge singularities
\begin{eqnarray}
\omega_{0 G G^{\perp}}^{(+)} &= \left ({4 N \over C_2 } \right )^{1/2}
\omega_{0 F \overline F}^{(+)} ,  \cr
\omega_{0 F \overline F}^{(+)} &= \left ( {g^2 C_2 \over 2 \pi^2 }
\right )^{1/2}.
\end{eqnarray}
The large $\omega_{0 G G^{\perp}} $
 is the origin of the large effect of the double logs on the small
$x$ behaviour in the flavour singlet $g_1 (x, Q^2)$ obtained in
\cite{BER96}.

This large effect demands to study the next-to leading log
contributions.
On the other hand it provides  a qualitative understanding of
the large gluon contribution to $g_1 (x, Q^2)$ which is observed in the
EMC effect.

{\large \bf Acknowledgements}

I thank the organizers of this workshop for invitation. I am grateful to
J. Bl\"umlein, L.N. Lipatov, L. Mankiewicz and L. Szymanowski for
discussions.

\end{document}